\documentclass[aps,prb,preprint,showpacs,superscriptaddress,floatfix]{revtex4}  

\usepackage{epsfig,amsmath,amssymb} 

\usepackage{graphics}
\usepackage{subfigure}

\begin{document}
\title
{
Effect of anisotropy on the ground-state magnetic ordering of the  \\
spin-half quantum $J_{1}^{XXZ}$--$J_{2}^{XXZ}$ model on the square lattice
}
\author
{
R.F.~Bishop
}
\affiliation
{
School of Physics and Astronomy, Schuster Building, The University of Manchester, Manchester, M13 9PL, UK}
\affiliation
{
School of Physics and Astronomy, University of Minnesota, 116 Church Street SE, Minneapolis, Minnesota 55455, USA}
\author
{
P.H.Y.~Li
}
\affiliation
{
School of Physics and Astronomy, Schuster Building, The University of Manchester, Manchester, M13 9PL, UK}
\affiliation
{
School of Physics and Astronomy, University of Minnesota, 116 Church Street SE, Minneapolis, Minnesota 55455, USA}

\author
{
R.~Darradi
}
\affiliation
{
Institut f\"ur Theoretische Physik, Universit\"at Magdeburg, P.O. Box 4120, 39016 Magdeburg, Germany}
\author
{
J.~Schulenburg
}
\affiliation
{
Universit\"{a}tsrechenzentrum, Universit\"{a}t Magdeburg, P.O. Box 4120, 39016 Magdeburg, Germany
}
\author
{
J.~Richter
}
\affiliation
{
Institut f\"ur Theoretische Physik, Universit\"at Magdeburg, P.O. Box 4120, 39016 Magdeburg, Germany}

\begin{abstract}
We study the zero-temperature phase diagram of the 2D quantum $J_{1}^{XXZ}$--$J_{2}^{XXZ}$
spin-$1/2$ anisotropic Heisenberg model on the square lattice.   In particular, the effects
of the anisotropy $\Delta$ on the $z$-aligned N\'{e}el and (collinear) stripe states, as well 
as on the $xy$-planar-aligned N\'{e}el and collinear stripe states, are examined.  All four of these 
quasiclassical states are chosen in turn as model states on top of which 
we systematically include the quantum correlations using a coupled cluster method analysis 
carried out to very high orders.  We find strong evidence for two {\it quantum triple points} 
(QTP's) at ($\Delta ^{c} = -0.10 \pm 0.15, J_{2}^{c}/J_{1} = 0.505 \pm 0.015$) and 
($\Delta ^{c} = 2.05 \pm 0.15, J_{2}^{c}/J_{1} = 0.530 \pm 0.015$), between which an 
intermediate magnetically-disordered phase emerges to separate the quasiclassical 
N\'{e}el and stripe collinear phases.  Above the upper QTP ($\Delta \gtrsim 2.0$) we find a 
direct first-order phase transition between the N\'{e}el and stripe phases, exactly 
as for the classical case.  The $z$-aligned and $xy$-planar-aligned phases meet precisely 
at $\Delta = 1$, also as for the classical case.  For all values of the anisotropy parameter between 
those of the two QTP's there exists a narrow range of values of $J_{2}/J_{1}$, 
$\alpha^{c_1}(\Delta)<J_{2}/J_{1}<\alpha^{c_2}(\Delta)$, centered near the point of maximum classical frustration, 
$J_{2}/J_{1} = \frac{1}{2}$, for which the intermediate phase exists.  This range is widest 
precisely at the isotropic point, $\Delta = 1$, where $\alpha^{c_1}(1) = 0.44 \pm 0.01$ and 
$\alpha^{c_2}(1) = 0.59 \pm 0.01$.  The two QTP's are characterized by values $\Delta = \Delta^{c}$ at 
which $\alpha^{c_1}(\Delta^{c}) = \alpha^{c_2}(\Delta^{c})$.
\end{abstract}

\maketitle

\section{Introduction}
\label{introd}
The exchange interactions that lead to collective magnetic behavior are 
clearly of purely quantum-mechanical origin.  Nevertheless, the underlying 
quantum nature has often safely been ignored in describing, at least at the 
qualitative level, many magnetic phenomena of interest in the past.  On the other 
hand, the investigation of magnetic systems and magnetic phenomena where the 
intrinsically quantal effects play a {\it dominant} role, and hence have to be 
accounted for in detail, has evolved in recent years to become a burgeoning 
area at the forefront of condensed matter theory.  Thus, the investigation of 
quantum magnets and their phase transitions, both quantum and thermal, has 
developed into an extremely active area of research.

From the experimental viewpoint major impetus has come both from the 
discovery of high-temperature superconductors and, since then, from the 
ever-increasing ability of materials scientists to fabricate a by now 
bewildering array of novel magnetic systems of reduced dimensionality, 
which display interesting quantum phenomena.\cite{Scholl:2004}  
While high-temperature superconductivity has raised the question of the link 
between the mechanism of superconductivity in the cuprates, for example, and 
spin fluctuations and magnetic order in one-dimensional (1D) 
and two-dimensional (2D) spin-half antiferromagnets, 
the new magnetic materials exhibit a wealth of new quantum phenomena of enormous 
interest in their own right.  

For example, in 1D systems, 
the universal paradigm of Tomonaga-Luttinger liquid~\cite{To:1950,Lu:1963} 
behavior has occupied a key position of interest, since Fermi liquid theory 
breaks down in 1D.  More generally, in {\it all} restricted geometries the 
interplay between reduced dimensionality, competing interactions and strong 
quantum fluctuations, generates a plethora of new states of condensed matter 
beyond the usual states of quasiclassical long-range order (LRO).  Thus, for 
high-temperature superconductivity, for example, it is suggested\cite{And:1987} 
that quantum spin fluctuation and frustration due to doping could lead to the 
collapse of the 2D N\'{e}el-ordered antiferromagnetic phase present at zero 
doping, and that this could be the clue for the superconducting behavior.  
This, and many similar experimental observations for other magnetic materials of 
reduced dimensionality, has intensified the study of order-disorder quantum 
phase transitions.  Thus, low-dimensional quantum antiferromagnets have attracted 
much recent attention as model systems in which strong quantum fluctuations 
might be able to destroy magnetic LRO in the ground state (GS).  
In the present paper we consider a system of $N \rightarrow \infty$ spin-1/2 
particles on a spatially isotropic 2D square lattice.

The spin-1/2 Heisenberg antiferromagnet with only nearest-neighbor (NN) bonds, 
all of equal strength, exhibits magnetic LRO at zero temperature on such 
bipartite lattices as the square lattice considered here.  A key mechanism 
that can then destroy the LRO for such systems, with a given lattice and spins of 
a given spin quantum number $s$, is the introduction of competing or 
frustrating bonds on top of the NN bonds.  The interested reader is 
referred to Refs.~[\onlinecite{Scholl:2004,Ma:1991}] for a more detailed 
discussion of 2D spin systems in general.

An archetypal model of the above type that has attracted much theoretical 
attention in recent years (see, e.g., 
Refs.~[\onlinecite{Ch:1988,Dag:1989,Ch:1990,Schulz:1992,Ri:1993,Ig:1993,Bi:1998_PRB58,Ca:2000,Ca:2001,Siu:2001,Su:2001,Be:2002,Sin:2003,Zhang:2003}])  
is the 2D spin-1/2 $J_{1}$--$J_{2}$ model on a square lattice with both 
NN and next-nearest-neighbor (NNN) antiferromagnetic interactions, with 
strength $J_{1} > 0$ and $J_{2} > 0$ respectively.  The NN bonds $J_{1} > 0$ 
promote N\'{e}el antiferromagnetic order, while the NNN bonds $J_{2} > 0$ act to frustrate 
or compete with this order.  All such frustrated quantum magnets continue to be 
of great theoretical interest because of the possible spin-liquid and other such 
novel magnetically disordered phases that they can exhibit (and see, e.g., Ref.\ [\onlinecite{Mi:2005}]).  
The recent syntheses of magnetic materials that can be well described by the spin-1/2 
$J_{1}$--$J_{2}$ model on the 2D square lattice, such as the undoped precursors to 
the high-temperature superconducting cuprates for small $J_{2}/J_{1}$ values, 
VOMoO$_{4}$ for intermediate $J_{2}/J_{1}$ values,\cite{Car:2002} and 
Li$_{2}$VOSiO$_{4}$ for large $J_{2}/J_{1}$ values,\cite{Mel:2000,Ro:2002} 
has fuelled further theoretical interest in the model.

The properties of the spin-1/2 $J_{1}$--$J_{2}$ model on the 2D square lattice are 
well understood in the limits when $J_{2}=0$ or $J_{1}=0$.  For the case when $J_{2}=0$, 
and the classical GS is perfectly N\'{e}el-ordered, the quantum fluctuations 
are not sufficiently strong enough to destroy the N\'{e}el LRO, although the staggered magnetization 
is reduced to about 61\% of its classical value.  Indeed, the best estimates for this 
order parameter are $61.4 \pm 0.1$\% from quantum Monte Carlo studies,\cite{Sa:1997} 
63.5\% from exact diagonalizations of small clusters,\cite{Ri:2004} $61.4 \pm 0.2$\% 
from series expansions,\cite{Zh:1991} $61.5 \pm 0.5$\% from the coupled cluster 
method (CCM) employed here,\cite{Bi:2000,Ri:2007,Bi:2008_IJMPB} and 61.4\% from third-order 
spin-wave theory.\cite{Ha:1992}  Clearly, they all agree remarkably well in this 
$J_{2}=0$ limit.  The opposite limit of large $J_{2}$ is a classic 
example\cite{Ch:1990} of the phenomenon of order by disorder.\cite{Vi:1977,Shen:1982}  
Thus, in the case where $J_{1} \rightarrow 0$ with $J_{2} \neq 0$ and fixed, the 
two sublattices each order antiferromagnetically at the classical level, but in 
directions which are independent of each other.  This degeneracy is lifted by quantum 
fluctuations and the GS becomes magnetically ordered collinearly as a stripe phase 
consisting of successive alternating rows (or columns) of parallel spins.

For intermediate values of $J_{2}/J_{1}$ it is now widely accepted that the quantum 
spin-1/2 $J_{1}$--$J_{2}$ model on the 2D square lattice has a ground-state (gs) phase 
diagram showing the above two phases with quasiclassical LRO (viz., a 
N\'{e}el-ordered ($\pi,\pi$) phase at smaller values of $J_{2}/J_{1}$, and a 
collinear stripe-ordered phase of the columnar ($\pi,0$) or row ($0,\pi$) type at 
larger values of $J_{2}/J_{1}$), separated by an intermediate quantum paramagnetic 
phase without magnetic LRO in the parameter regime 
$\alpha^{c_1}<J_{2}/J_{1}<\alpha^{c_2}$, where $\alpha^{c_1}\approx 0.4$ and 
$\alpha^{c_2}\approx 0.6$.  The precise nature of the intermediate magnetically-disordered 
phase  is still not fully resolved.  Suggested candidates include a homogeneous 
spin-liquid state of various types with no broken 
symmetry (see, e.g., Ref.\ [{\onlinecite{Zhang:2003}]), or a valence-bond solid (VBS) phase with some 
broken symmetry.  Possible spin-liquid states include a resonating-valence-bond (RVB) 
state proposed by Anderson,\cite{And:1987} which has been supported more recently by 
variational quantum Monte Carlo studies.\cite{Ca:2001}  Other studies 
\cite{Dag:1989,Read:1991,Singh:1999,Ko:2000,Sir:2006} have supported a spontaneously dimerized 
state for the intermediate phase with both translational and rotational 
symmetry broken, and thus representing a columnar VBS phase.  Yet other 
studies\cite{Ca:2000,Zh:1996} have supported instead a plaquette VBS state for the 
intermediate phase, with translational symmetry broken but with rotational 
symmetry preserved.

There has also been considerable discussion in recent years as to whether the 
quantum phase transition between the quasiclassical N\'{e}el phase and the 
magnetically disordered (intermediate paramagnetic) phase in the spin-1/2 
$J_{1}$--$J_{2}$ model on the 2D square lattice is first-order or of continuous 
second-order type.  A particularly intriguing suggestion by Senthil {\it et al.}\cite{Sen:2004} 
is that there is a second-order phase transition in the model between 
the N\'{e}el state and the intermediate disordered state (which these authors argue is 
a VBS state), which is not described by a Ginzburg-Landau-type critical theory, but 
is rather described in terms of a deconfined quantum critical point.  Such direct 
second-order quantum phase transitions between two states with different broken 
symmetries, and which are hence characterized by two seemingly independent order 
parameters, are difficult to understand within the standard critical theory 
approach of Ginzburg and Landau, as we indicate below.

Thus, the competition between two such distinct kinds of quantum order associated 
with different broken symmetries would lead generically in the Ginzburg-Landau 
scenario to one of only three possibilities: (i) a first-order transition between the two 
states, (ii) an intermediate region of co-existence between both phases with both kinds
of order present, or (iii) a region of intermediate phase with neither of the orders of 
these two phases present.  A direct second-order transition between states of different 
broken symmetries is only permissible within the standard Ginzburg-Landau critical 
theory if it arises by an accidental fine-tuning of the disparate order parameters to 
a multicritical point.  Thus, for the spin-1/2 $J_{1}$--$J_{2}$ model on 
the 2D square lattice and its quantum phase transition suggested by 
Senthil {\it et al.},\cite{Sen:2004} it would require the completely accidental 
coincidence (or near coincidence) of the point where the magnetic order parameter (i.e., 
the staggered magnetization) vanishes for the N\'{e}el phase with the point where the 
dimer order parameter vanishes for the VBS phase.  Since each of these phases has 
a different broken symmetry (viz., spin-rotation symmetry for the N\'{e}el phase and the 
lattice symmetry for the VBS phase), one would naively expect that each transition 
is described by its own independent order parameter (i.e., the staggered magnetization 
for the N\'{e}el phase and the dimer order parameter for the VBS phase) and that the two 
transitions should hence be mutually independent.  

By contrast, the ``deconfined'' type of quantum phase transition postulated by 
Senthil {\it et al.}\cite{Sen:2004} permits direct second-order quantum phase transitions 
between such states with different forms of broken symmetry.  In their scenario the 
quantum critical points still separate phases characterized by order parameters of the 
conventional (i.e., in their language, ``confining'') kind, but their proposed new critical 
theory involves fractional degrees of freedom (viz., spinons for the spin-1/2 $J_{1}$--$J_{2}$ 
model on the 2D square lattice) that interact via an emergent gauge field.  
For our specific example the order parameters of both the N\'{e}el and VBS phases discussed 
above are represented in terms of the spinons, which themselves become ``deconfined'' 
exactly at the critical point.  The postulate that the spinons are the fundamental constituents 
of both order parameters then affords a natural explanation for the direct second-order phase 
transition between two states of the system that otherwise seem very different on the 
basis of their broken symmetries. 

We note, however, that the deconfined phase transition theory of Senthil 
{\it et al.}\cite{Sen:2004} is still the subject of controversy.  Other authors believe 
that the phase transition in the  spin-1/2 $J_{1}$--$J_{2}$ model on the 2D square lattice from 
the Ne\'{e}l phase to the intermediate magnetically-disordered phase need not be due to 
a deconfinement of spinons.  For example, Sirker {\it et al.}\cite{Sir:2006} have 
argued on the basis of both spin-wave theory and numerical results from series 
expansion analyses, that this transition is more likely to be a (weakly) first-order 
transition between the Ne\'{e}l phase and a VBS phase with 
columnar dimerization.  Other authors have also proposed other, perhaps less radical, 
mechanisms to explain such second-order phase transitions (if they exist) and their seeming disagreement 
(except by accidental fine tuning) with Ginzburg-Landau theory.  What seems clearly to be 
a minimal requirement is that the order parameters of the two phases with different broken 
symmetry should be related in some way.  Thus, a Ginzburg-Landau-type theory can 
only be preserved if it contains additional terms in the effective theory that represent
interactions between the two order parameters.  For example, just such an effective 
theory has been proposed for the 2D spin-1/2 $J_{1}$--$J_{2}$ model on the square lattice by 
Sushkov {\it et al.},\cite{Su:2002} and further discussed by Sirker 
{\it et al.}\cite{Sir:2006}

From the classical viewpoint frustrated models often exhibit ``accidental'' degeneracy, 
and the degree of such degeneracy, which can vary enormously, has become widely viewed 
as a measure of the frustration.  Among the effects that can act to lift any such 
degeneracy are thermal fluctuations, quantum fluctuations, and such ``perturbations'' as 
spin-orbit interactions, spin-lattice couplings, further neglected exchange terms, and 
impurities, all of which might be present in actual materials.  In the present paper we 
focus particular attention on the role of quantum fluctuations.  From the quantum viewpoint 
such frustrated quantum magnets as the spin-1/2 $J_{1}$--$J_{2}$ model on the 2D square 
lattice often have ground states that are macroscopically degenerate.  This feature
leads naturally to an increased sensitivity of the underlying Hamiltonian to the presence 
of small perturbations.  In particular, the presence in real systems that are well 
characterised by the $J_{1}$--$J_{2}$ model, of anisotropies, either in spin space or 
in real space, naturally raises the issue of how robust are the properties of the model 
against any such perturbations. 

Combining the above two viewpoints, it is clear that it is of particular interest in the 
study of frustrated quantum magnets to focus special attention on the mechanisms or 
parameters that are available to us to ``tune'' or vary the quantum fluctuations that 
play such a key role in determining their gs phase structures.  Apart from changing the 
spin quantum number or the dimensionality and lattice type of the system, or tuning the 
relative strengths of the competing exchange interactions, another key mechanism 
is the introduction of anisotropy into the existing exchange bonds.  Such anisotropy 
can be either in real space\cite{Ne:2003,Si:2004,Star:2004,Mo:2006,Bi:2007_j1j2j3_spinHalf,Bi:2008} 
or in spin space.\cite{Ro:2004,Via:2007,Be:1998,Da:2004} 

In order to investigate the effect in real space an interesting generalization of the 
pure $J_{1}$--$J_{2}$ model has been introduced recently by Nersesyan and 
Tsvelik\cite{Ne:2003} and further studied by other groups including 
ourselves.\cite{Si:2004,Star:2004,Mo:2006,Bi:2007_j1j2j3_spinHalf,Bi:2008} 
This generalization, the so-called $J_{1}$--$J_{1}'$--$J_{2}$ model, introduces a spatial 
anisotropy into the 2D $J_{1}$--$J_{2}$ model on the square lattice by allowing the NN bonds 
to have different strengths $J_{1}$ and $J_{1}'$ in the two orthogonal spatial lattice dimensions, 
while keeping all of the NNN bonds across the diagonals to have the same strength $J_{2}$.  
In previous work of our own\cite{Bi:2007_j1j2j3_spinHalf,Bi:2008} on this $J_{1}$--$J_{1}'$--$J_{2}$ 
model we studied the effect of the coupling $J_{1}'$ on the quasiclassical N\'{e}el-ordered and 
stripe-ordered phases for both the spin-1/2 and spin-1 cases.  For the spin-1/2 
case,\cite{Bi:2007_j1j2j3_spinHalf} we found the surprising and novel result that 
there exists a quantum triple point below which there is a second-order phase transition 
between the quasiclassical N\'{e}el and columnar stripe-ordered phases with magnetic
LRO, whereas only above this point are these two phases separated by the intermediate 
magnetically disordered phase seen in the pure spin-1/2 $J_{1}$--$J_{2}$ model on the 2D 
square lattice (i.e., at $J_{1}' = J_{1}$).  We found that the quantum critical points for both of 
the quasiclassical phases with magnetic LRO increase as the coupling ratio $J_{1}'/J_{1}$ is 
increased, and an intermediate phase with no magnetic LRO emerges only when 
$J_{1}'/J_{1} \gtrsim 0.6$, with strong indications of a quantum triple point at 
$J_{1}'/J_{1} = 0.60 \pm 0.03, J_{2}/J_{1} = 0.33 \pm 0.02$.  
For $J_{1}'/J_{1}=1$, the results agree with the previously known results of 
the $J_{1}$--$J_{2}$ model described above.

In the present paper we generalize the spin-1/2 $J_{1}$--$J_{2}$ model on the 2D 
square lattice in a different direction by allowing the bonds to become anisotropic 
in spin space rather than in real space.  Such spin anisotropy is relevant 
experimentally as well as theoretically, since it is likely to be present, if only weakly, 
in any real material.  Furthermore, the intermediate magnetically-disordered phase  
is likely to be particularly sensitive to {\it any} tuning of the quantum 
fluctuations, as we have seen above in the case of spatial anisotropy.  Indeed, other 
evidence indicates that the intermediate phase might even disappear altogether in certain 
situations, such as increasing the dimensionality or the spin quantum number.  

Thus, for example, the influence of frustration and quantum fluctuations 
on the magnetic ordering in the GS of the spin-1/2 $J_{1}$--$J_{2}$ model 
on the body-centered cubic (bcc) lattice has been studied using exact 
diagonalization of small lattices and linear spin-wave theory,\cite{Sch:2002} 
and also by using linked-cluster series expansions.\cite{Oi:2004}  
Contrary to the results for the corresponding model on the square
lattice, it was found for the bcc lattice that frustration and quantum fluctuations 
do not lead to a quantum disordered phase for strong frustration.  Rather, the results of all  
approaches suggest a first-order quantum phase transition at a value $J_{2}/J_{1} \approx 0.70$
from the quasiclassical N\'{e}el phase at low $J_2$ to a quasiclassical collinear phase 
at large $J_2$.  Similarly, the intermediate phase can also disappear when the spin 
quantum number $s$ is increased for the $J_{1}$--$J_{2}$ model on the 2D square lattice.  Thus, 
we\cite{Bi:2008} found no evidence for a magnetically disordered state (for larger values of 
$J_2/J_1$) for the $s=1$ case, by contrast with the $s=1/2$) case.\cite{Bi:2007_j1j2j3_spinHalf} 
Instead, we found a quantum tricritical point in the $s=1$ case of the $J_{1}$--$J_{1}'$--$J_{2}$ 
model on the 2D square lattice at $J_{1}'/J_{1} = 0.66 \pm 0.03, J_{2}/J_{1} = 0.35 \pm 0.02$, 
where a line of second-order phase transitions between the quasiclassical N\'{e}el and columar 
stripe-ordered phases (for $J_{1}'/J_{1} \lesssim 0.66$) meets a line of first-order 
phase transitions between the same two phases (for $J_{1}'/J_{1} \gtrsim 0.66$).

As in our previous work\cite{Bi:2007_j1j2j3_spinHalf,Bi:2008} involving 
the effect of spatial anisotropy on the spin-1/2 and spin-1 $J_{1}$--$J_{2}$ models 
on the 2D square lattice, we again employ the coupled cluster method (CCM) 
to investigate now the effect on the same model of spin anisotropy.  
The CCM is one of the most powerful techniques in microscopic quantum many-body 
theory.\cite{Bi:1991,Bi:1998}  It has been applied successfully to many quantum 
magnets.\cite{Fa:2004,Ze:1998,Kr:2000,Bi:2000,Fa:2002,Dar:2005,Schm:2006}    
It is capable of calculating with high accuracy the ground- and excited-state properties of 
spin systems.  In particular, it is an effective tool for studying highly frustrated 
quantum magnets, where such other numerical methods as the quantum Monte Carlo method and 
the exact diagonalization method are often severely limited in practice, e.g., by the 
``minus-sign problem'' and the very small sizes of the spin systems that can be 
handled in practice with available computing resources, respectively.

\section{The model}
The usual 2D spin-1/2 $J_{1}$--$J_{2}$ model is an isotropic Heisenberg model on a 
square lattice with two kinds of exchange bonds, with strength $J_{1}$ for the 
NN bonds along both the row and the column directions, and with strength 
$J_{2}$ for the NNN bonds along the diagonals, as shown in Fig.\ \ref{model}(a). 
\begin{figure*}
\scalebox{0.6}{
 \epsfig{file=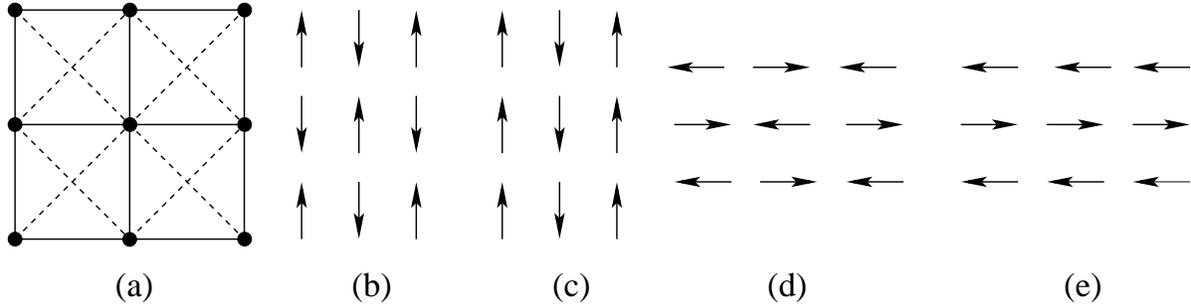}}
\caption{(a) The $J_{1}^{XXZ}$--$J_{2}^{XXZ}$ Heisenberg model; --- $J_{1}$; - - - $J_{2}$; 
(b) and (c) $z$-aligned states for the N\'{e}el and stripe columnar phases respectively; 
(d) and (e) planar $x$-aligned states for the N\'{e}el and stripe columnar phases respectively.  
Arrows in (b), (c), (d) and (e) represent spins situated on the sites of the square 
lattice [symbolized by \textbullet \; in (a)].}
\label{model}
\end{figure*}
Here we generalize the model by including an anisotropy in spin space in both 
the NN and NNN bonds.  We are aware of only a very few earlier investigations with a similar 
goal.\cite{Ro:2004,Via:2007,Be:1998}  The two most detailed 
have studied the extreme limits where either the frustrating 
NNN interaction becomes anisotropic but the NN interaction remains isotropic\cite{Ro:2004} 
(viz., the $J_{1}$--$J_{2}^{XXZ}$ model) and the opposite case where the NN 
interaction becomes anisotropic but the NNN interaction remains isotropic\cite{Via:2007} 
(viz., the $J_{1}^{XXZ}$--$J_{2}$ model).  In real materials one might expect 
both exchange interactions to become anisotropic.  To our knowledge the only study of 
this case\cite{Be:1998} (viz., the $J_{1}^{XXZ}$--$J_{2}^{XXZ}$ model) has 
been done using the rather crude tool of linear spin-wave 
theory (LSWT), from which it is notoriously difficult to draw any firm quantitative conclusions 
about the positions of the gs phase boundaries of a system.  It is equally difficult 
to use LSWT to predict with confidence either the number of phases present in the gs phase 
diagram  or the nature of the quantum phase transitions between them.  
We comment further on the application of spin-wave theory to the $J_1$--$J_2$ model 
and its generalizations in Sec.\ \ref{discussion}.  The aim of the 
present paper is to use the CCM, as a much more accurate many-body tool, to investigate the 
spin-1/2 $J_{1}^{XXZ}$--$J_{2}^{XXZ}$ model on the 2D square lattice.  

In order to keep the size of the parameter space manageable the anisotropy parameter 
$\Delta$ is assumed to be the same in both exchange terms, 
thus yielding the so-called $J_{1}^{XXZ}$--$J_{2}^{XXZ}$ model, whose Hamiltonian is 
described by 
\begin{eqnarray}
H &=& J_{1}\sum_{\langle i,j \rangle}(s^{x}_{i}s^{x}_{j}+s^{y}_{i}s^{y}_{j}+\Delta s^{z}_{i}s^{z}_{j}) \nonumber \\
 & + & J_{2}\sum_{\langle\langle i,k \rangle\rangle}(s^{x}_{i}s^{x}_{k}+s^{y}_{i}s^{y}_{k}+\Delta s^{z}_{i}s^{z}_{k})\,,  \label{H}
\end{eqnarray}
where the sums over $\langle i,j \rangle$ and $\langle\langle i,k \rangle\rangle$ 
run over all NN and NNN pairs respectively, counting each bond once and once only. 
We are interested only in the case of competing antiferromagnetic bonds, 
$J_{1} > 0$ and $J_{2} > 0$, and henceforth, for all of the results shown in 
Sec.\ \ref{results}, we set $J_{1}=1$.  Similarly, we shall be interested essentially 
only in the region $\Delta >0$ (although for reasons discussed below in 
Sec.\ \ref{results} we shall show results also for small negative values of $\Delta$).

This model has two types of classical antiferromagnetic ground states, namely a 
$z$-aligned state for $\Delta > 1$ and an $xy$-planar-aligned state for 
$0 < \Delta < 1$.  Since all directions in the $xy$-plane in spin space 
are equivalent, we may choose the direction arbitrarily to be the $x$-direction, 
say.  Both of these $z$-aligned and $x$-aligned states further divide into a 
N\'{e}el ($\pi,\pi$) state and stripe states (columnar stripe
($\pi,0$) and row stripe ($0,\pi$)), the spin orientations of which
are shown in Figs.~\ref{model}(b,c,d,e) accordingly.  There is clearly a symmetry
under the interchange of rows and columns, which implies that we need only consider the 
columnar stripe states.  The (first-order) classical phase transition occurs 
at $J^{c}_{2}=\frac{1}{2}J_{1}$, with the N\'{e}el states being the classical GS for 
$J_{2} < \frac{1}{2}J_{1}$, and the columnar stripe states being the classical 
GS for $J_{2} > \frac{1}{2}J_{1}$.

\section{The coupled cluster method}
We briefly outline the CCM formalism (and see
Refs.~[\onlinecite{Bi:1991,Bi:1998,Ze:1998,Bi:2000,Kr:2000,Fa:2002,Fa:2004,Dar:2005,Schm:2006}]
for further details).  The first step of any CCM calculation is to choose 
a normalized model (or reference) state $|\Phi\rangle$ which can act as a cyclic 
vector with respect to a complete set of mutually commuting multi-configurational 
creation operators, $C^{+}_{I} \equiv (C^{-}_{I})^{\dagger}$.  The index $I$ here 
is a set-index that labels the many-particle configuration created in the 
state $C^{+}_{I}|\Phi\rangle$.  The requirements are that any many-particle state 
can be written exactly and uniquely as a linear combination of the states 
$\{C^{+}_{I}|\Phi\rangle\}$, together with the conditions,
\begin{equation}
\langle\Phi| C^{+}_{I} = 0 = C^{-}_{I}|\Phi\rangle \quad \forall \emph{I} \neq 0\,; \quad C^{+}_{0} \equiv 1 \, ,
\end{equation}
\begin{equation}
[C^{+}_{I},C^{+}_{J}] = 0 = [C^{-}_{I},C^{-}_{J}]\,.  \label{C_comm}
\end{equation}

The Schr\"{o}dinger equations for the many-body ground-state (gs) ket and bra states are
\begin{subequations}
\begin{eqnarray}
H|\Psi\rangle &= E|\Psi\rangle \,,    \label{Sch_eq_ket} \\      
\langle\tilde{\Psi}|H &=E\langle\tilde{\Psi}|\,  \label{Sch_eq_bra}\,,
\end{eqnarray}
\end{subequations}
respectively, with normalization chosen such that
$\langle\tilde{\Psi}|\Psi\rangle = 1$ 
[i.e., with $\langle\tilde{\Psi}|=(\langle\Psi|\Psi\rangle)^{-1}\langle\Psi|$], 
and with $|\Psi\rangle $ itself satisfying the intermediate normalization 
condition $\langle\Phi|\Psi\rangle=1=\langle\Phi|\Phi\rangle$.  In terms 
of the set $\{|\Phi\rangle; C^{+}_{I}\}$, the CCM employs the exponential 
parametrization
\begin{subequations}
\begin{eqnarray}
|\Psi\rangle &= \mbox{e}^{S}|\Phi\rangle\,, \quad S & = \sum_{I\neq0}{\cal S}_{I}C^{+}_{I}  \label{ccm_para_ket}
\end{eqnarray}
for the exact gs ket energy eigenstate.  Its counterpart for the exact 
gs bra energy eigenstate is chosen as
\begin{eqnarray}
\langle\tilde{\Psi}|=\langle\Phi|\tilde{S}\mbox{e}^{S}\,, \quad  \tilde{S} & = 1 + \sum_{I\neq0}\tilde{{\cal S}_{I}}C^{-}_{I}\,.    \label{ccm_para_bra}
\end{eqnarray}
\end{subequations}

It is important to note that while the parametrizations of Eqs.\ (\ref{ccm_para_ket}) 
and (\ref{ccm_para_bra}) are not manifestly Hermitian-conjugate, they do preserve 
the important Hellmann-Feynman theorem at all levels of approximation (viz., 
when the complete set of many-particle configurations $\{I\}$ is truncated).\cite{Bi:1998}  
Furthermore the amplitudes $({\cal S}_{I}, \tilde{{\cal S}_{I}})$ form canonically 
conjugate pairs in a time-dependent version of the CCM, by contrast with the 
pairs $({\cal S}_{I}, {\cal S}_{I}^{\ast})$, coming from a manifestly 
Hermitian-conjugate representation for 
$\langle\tilde{\Psi}|=(\langle\Phi|\mbox{e}^{S^{\dagger}}\mbox{e}^{S}|\Phi)^{-1}\langle\Phi|\mbox{e}^{S^{\dagger}}$, 
that are {\it not} canonically conjugate to one another.\cite{Bi:1998}

The static gs CCM correlation operators, $S$ and $\tilde{S}$, contain 
the real c-number correlation coefficients, ${\cal S}_{I}$ and $\tilde{{\cal S}_{I}}$,  
that need to be calculated.  Clearly, once the coefficients 
$\{{\cal S}_{I}, \tilde{{\cal S}_{I}}\}$ are known, all other gs properties 
of the many-body system can be derived from them.  To find the gs correlation 
coefficients we simply insert the parametrizations of Eqs.\ (\ref{ccm_para_ket},b) into 
the Schr\"{o}dinger equations (\ref{Sch_eq_ket},b) and project onto the complete 
sets of states $\langle\Phi|C^{-}_{I}$ and $C^{+}_{I}|\Phi\rangle$, respectively.  
Completely equivalently, we may simply demand that the gs energy expectation 
value, $\bar{H} \equiv \langle\tilde{\Psi}|H|\Psi\rangle$, is minimized with respect 
to the entire set $\{{\cal S}_{I}, \tilde{{\cal S}_{I}}\}$.  In either case 
we are easily led to the equations
\begin{subequations}
\begin{eqnarray}
\langle \Phi|C^{-}_{I}\mbox{e}^{-S}H\mbox{e}^{S}|\Phi\rangle & = 0\;; \quad  \forall I \neq 0\,,   \label{ket_coeff} \\
\langle\Phi|\tilde{S}\mbox{e}^{-S}[H, C^{+}_{I}]\mbox{e}^{S}|\Phi\rangle & = 0\;; \quad \forall I \neq 0 \,,  \label{bra_coeff}
\end{eqnarray}
\end{subequations}
which we then solve for the set $\{{\cal S}_{I}, \tilde{{\cal S}_{I}}\}$.  
Equation (\ref{ket_coeff}) also shows that the gs energy at the stationary point has 
the simple form
\begin{equation}
E = E(\{{\cal S}_{I}\})=\langle\Phi|\mbox{e}^{-S}H\mbox{e}^{S}|\Phi\rangle\,.   \label{Energy}
\end{equation}
It is important to realize that this (bi-)variational formulation does not 
necessarily lead to an upper bound for $E$ when the summations for $S$ and 
$\tilde{S}$ in Eqs.\ (\ref{ccm_para_ket},b) are truncated, due to the lack of 
manifest Hermiticity when such approximations are made.  Nonetheless, 
one can prove\cite{Bi:1998} that the important Hellmann-Feynman 
theorem {\it is} preserved in all such approximations.

We note that Eq.\ (\ref{ket_coeff}) represents a coupled set of non-linear 
multinomial equations for the c-number correlation coefficients $\{ {\cal S}_{I} \}$.  
The nested commutator expansion of the similarity-transformed Hamiltonian,
\begin{equation}
\mbox{e}^{-S}H\mbox{e}^{S} = H + [H,S] + \frac{1}{2!}[[H,S],S] + \cdots \,,  \label{H_Sim_xform}
\end{equation}
and the fact that all of the individual components of $S$ in the expansion of 
Eq.\ (\ref{ccm_para_ket}) commute with one another by construction [and see Eq.\ (\ref{C_comm})], 
together imply that each element of $S$ in Eq.\ (\ref{ccm_para_ket}) is linked directly 
to the Hamiltonian in each of the terms in Eq.\ (\ref{H_Sim_xform}).  Thus, each of the 
coupled equations (\ref{ket_coeff}) is of Goldstone {\it linked-cluster} type.  
In turn, this guarantees that all extensive variables, such as the energy, 
scale linearly with particle number, $N$.  Thus, at any level of approximation 
obtained by truncation in the summations on the index $I$  in Eqs.\ (\ref{ccm_para_ket},b), 
we may (and do) always work from the outset in the limit $N \rightarrow \infty$ 
of an infinite system.

Furthermore, each of the linked-cluster equations (\ref{ket_coeff}) is of finite 
length when expanded, since the otherwise infinite series of Eq.\ (\ref{H_Sim_xform}) 
will always terminate at a finite order, provided only (as is usually the case, 
including that of the Hamiltonian considered here) that each term in the Hamiltonian, $H$, 
contains a finite number of single-particle destruction operators defined with 
respect to the reference (vacuum) state $|\Phi\rangle$.  Hence the CCM 
parametrization naturally leads to a workable scheme that can be computationally 
implemented in a very efficient manner.

Before discussing the possible CCM truncation schemes, we note that it is very 
convenient to treat the spins on each lattice site in a chosen model state $|\Phi\rangle$ 
as equivalent.  In order to do so we introduce a different local quantization 
axis and a correspondingly different set of 
spin coordinates on each site, so that all spins, whatever 
their original orientations in $|\Phi\rangle$ in a global spin-coordinate system, 
align along the negative $z$-direction, say, in these local spin coordinates.  
This can always be done by defining a suitable rotation in spin space of the 
global spin coordinates at each lattice site.  Such rotations are canonical 
transformations that leave the spin commutation relations unchanged.  In these 
local spin axes where the configuration indices $I$ simply become a set of lattice 
site indices, $I \rightarrow \{k_{1},k_{2},\cdots k_{m}\}$, the generalized 
multi-configurational creation operators $C^{+}_{I}$ are simple products of 
single spin-raising operators, 
$C^{+}_{I} \rightarrow s^{+}_{k_{1}} s^{+}_{k_{2}} \cdots s^{+}_{k_{m}}$, 
where $s^{\pm}_{k} \equiv s^{x}_{k} \pm is^{y}_{k}$\,, 
and $(s^{x}_{k}, s^{y}_{k}, s^{z}_{k})$ are the usual SU(2) spin operators 
on lattice site $k$.  For the quasiclassical 
magnetically-ordered states that we calculate here, the order parameter is the 
sublattice magnetization, $M$, which is given within our local spin coordinates 
defined above as
\begin{equation}
M \equiv -\frac{1}{N} \langle\tilde{\Psi}|\sum_{k=1}^{N}s^{z}_{k}|\Psi\rangle\,.
\end{equation}

It is usually convenient to take the classical ground states as 
our (initial) choices for the model state $|\Phi\rangle$.  Hence, 
we may choose here either a N\'{e}el state or a (columnar) 
stripe state for $|\Phi \rangle$.  Each of these can be further sub-divided into 
a $z$-aligned choice or a planar (say, $x$-aligned) choice, which we expect to be 
appropriate for $\Delta > 1$ and $|\Delta|<1$ respectively on purely classical 
grounds.  We present results below in Sec.\ \ref{results} based on all four of these 
classical ground states as choices for $|\Phi\rangle$.

Clearly the CCM formalism is exact when one includes all possible multi-spin 
configurations $I$ in the sums in Eqs.\ (\ref{ccm_para_ket},b) for the cluster 
correlation operators $S$ and $\tilde{S}$.  In practice, however, truncations are 
needed.  As in much of our previous work for spin-half models we employ here the 
so-called LSUB$n$ 
scheme,\cite{Bi:1991,Bi:1998,Ze:1998,Bi:2000,Kr:2000,Fa:2002,Fa:2004,Dar:2005,Schm:2006}  
in which all possible multi-spin-flip correlations over different locales on the 
lattice defined by $n$ or fewer contiguous lattice sites are retained.  (Two sites 
are defined to be contiguous here if they are NN sites on the lattice.)  The numbers 
of such fundamental configurations (viz., those that are distinct under the symmetries 
of the Hamiltonian and of the model state $|\Phi\rangle$) that are retained for the 
$z$-aligned and planar $x$-aligned states of the current model in their N\'{e}el and 
stripe phases in the various LSUB$n$ approximations are shown in Table~\ref{FundConf}.  
\begin{table}
\begin{center}
\caption{Numbers of fundamental configurations ($\sharp$ f.c.) retained in the 
CCM LSUB$n$ approximation for the $z$-aligned states and the planar $x$-aligned 
states of the $s=1/2$ $J_{1}^{XXZ}$--$J_{2}^{XXZ}$ model.}
\begin{tabular}{cccccc} \hline\hline
 & \multicolumn{2}{c}{$z$-aligned states}  &  & \multicolumn{2}{c}{planar $x$-aligned states} \\ 
\\[-7pt] \cline{1-3} \cline{5-6} \\[-6pt]
 Scheme  & \multicolumn{2}{c}{$\sharp$ f.c.}  &   &\multicolumn{2}{c}{$\sharp$ f.c.} \\
\\[-7pt] \cline{2-3} \cline{5-6} \\[-6pt]
& N\'{e}el  & stripe   &  & N\'{e}el & stripe \\ 
\\[-7pt] \cline{1-3}  \cline{4-6} \\[-6pt]
LSUB$2$ & 1 & 1  &   & 1 &  2 \\
LSUB$4$ &  7 & 9  &  & 10  & 18  \\
LSUB$6$ & 75 & 106 &  & 131  &  252 \\
LSUB$8$ &  1287 & 1922  &   & 2793  & 5532  \\
LSUB$10$ & 29605 & 45825  &    &  74206 &  148127 \\ \hline\hline
\end{tabular}
\label{FundConf}
\end{center}
\end{table}
Parallel computing is employed to solve the corresponding coupled 
sets of CCM bra- and ket-state equations (\ref{ket_coeff},b).\cite{ccm} 
Our computing power is such that we can obtain LSUB$n$ 
results for $n = \{2,4,6,8,10\}$ for both the $z$-aligned model states and the 
$x$-aligned model states, as shown in Table \ref{FundConf}.  However the very large numbers 
of fundamental configurations retained in the latter case at the LSUB10 level is only 
possible with supercomputing resources.  For example, the solution of the equations involving the 
nearly 150,000 fundamental configurations for the stripe phase of the planar $x$-aligned 
state required the simultaneous use of 600 processors running for approximately 6 hours, for 
each value of the anisotropy parameter $\Delta$ in the Hamiltonian of Eq.\ (\ref{H}).

The final step in any CCM calculation is then to extrapolate the approximate 
LSUB$n$ results to the exact, $n \rightarrow \infty$, limit.  Although no 
fundamental theory is known on how the LSUB$n$ data for such physical quantities 
as the gs energy per spin, $E/N$, and the gs staggered magnetization, $M$, scale 
with $n$ in the $n \rightarrow \infty$, limit, we have a great deal of experience 
in doing so from previous 
calculations.\cite{Ze:1998,Kr:2000,Bi:2000,Dar:2005,Schm:2006,Bi:2007_j1j2j3_spinHalf,Bi:2008,Ri:2007,Zi:2008}  
Thus, we employ here the same well-tested LSUB$n$ scaling laws as we have used, 
for example, for the $J_{1}$--$J_{1}'$--$J_{2}$ model,\cite{Bi:2007_j1j2j3_spinHalf,Bi:2008}
namely  
\begin{equation}
E/N=a_{0}+a_{1}n^{-2}+a_{2}n^{-4}  \label{ext_E}
\end{equation} 
for the gs energy per spin, and
\begin{equation}
M=b_{0}+n^{-0.5}\left(b_{1}+b_{2}n^{-1}\right)  \label{ext4}
\end{equation} 
for the gs staggered magnetization, both of which have been successfully used previously 
for systems showing an order-disorder quantum phase transition.  An alternative 
leading power-law extrapolation scheme for the order parameter,
\begin{equation}
M = c_{0}+c_{1}n^{-c_{2}}\,,  \label{ext1}
\end{equation}
has also been successfully used previously to determine the phase transition points.  
For most systems with order-disorder transitions the two extrapolation schemes of Eqs.\
(\ref{ext4}) and (\ref{ext1}) give remarkably similar results almost everywhere, 
as demonstrated explicitly, for example, for the case of quasi-one-dimensional quantum 
Heisenberg antiferromagnets with a weak interchain coupling.\cite{Zi:2008}  However, in regions 
very near quantum triple points the form of Eq.\ (\ref{ext4}) is more robust than that 
of Eq.\ (\ref{ext1}) due to the addition of the next-to-leading correction term, as has 
been explained in detail elsewhere.\cite{Bi:2007_j1j2j3_spinHalf}  Hence, in this 
work we use the extrapolation schemes of Eqs.\ (\ref{ext_E}) and (\ref{ext4}).

Obviously, better results are obtained from the LSUB$n$ extrapolation schemes if the 
data with the lowest $n$ values are not used in the fits.  However, a robust and stable 
fit to any fitting formula with $m$ unknown parameters is generally only obtained by using 
at least ($m+1$) data points.  In particular, a fit to 
only $m$ data points should be avoided whenever possible.  
In our case both fitting schemes in Eqs.\ (\ref{ext_E}) 
and (\ref{ext4}) have $m=3$ unknown parameters 
to be determined.  For all four model states we have LSUB$n$ data with $n=\{2,4,6,8,10\}$, and 
it is clear that the optimal fits should be obtained using the sets $n=\{4,6,8,10\}$.  All the 
extrapolated results that we present below in Sec.\ \ref{results} are 
obtained in precisely this way.  However, we have 
also extrapolated $E/N$ and $M$ using the sets $n=\{2,4,6,8,10\}$ and $n=\{4,6,8\}$.  In almost 
all cases they lead to very similar results, which adds credence to the stability of our numerical 
results and to the validity of our conclusions presented below.

\section{Results}
\label{results}
Figure~\ref{E}
\begin{figure*}
\begin{center}
\mbox{
  \subfigure[$z$-aligned states]{\label{E_Zaligned}\scalebox{0.3}{\epsfig{file=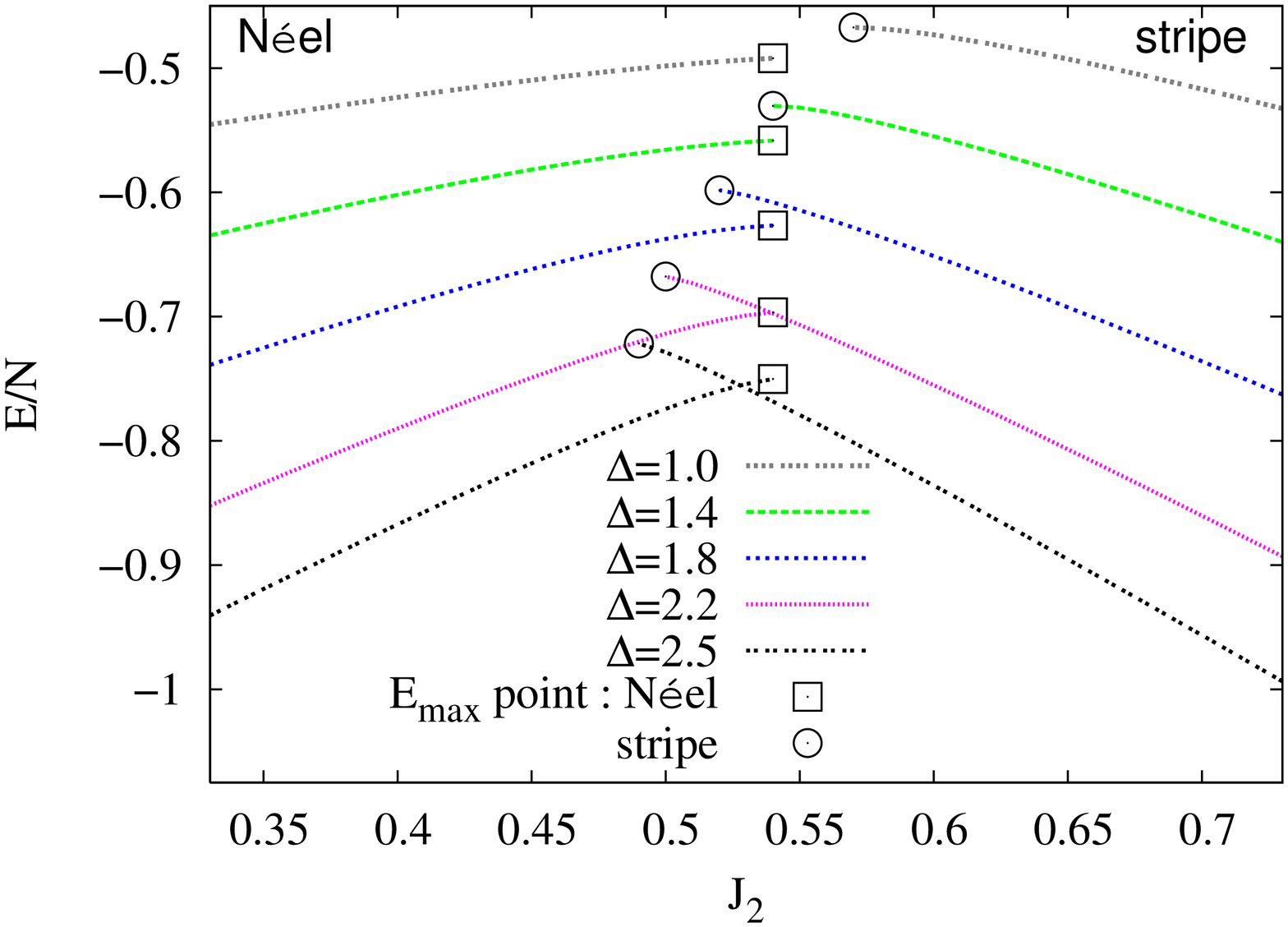,angle=270}}}
  \subfigure[planar $x$-aligned states]{\label{E_planar}\scalebox{0.3}{\epsfig{file=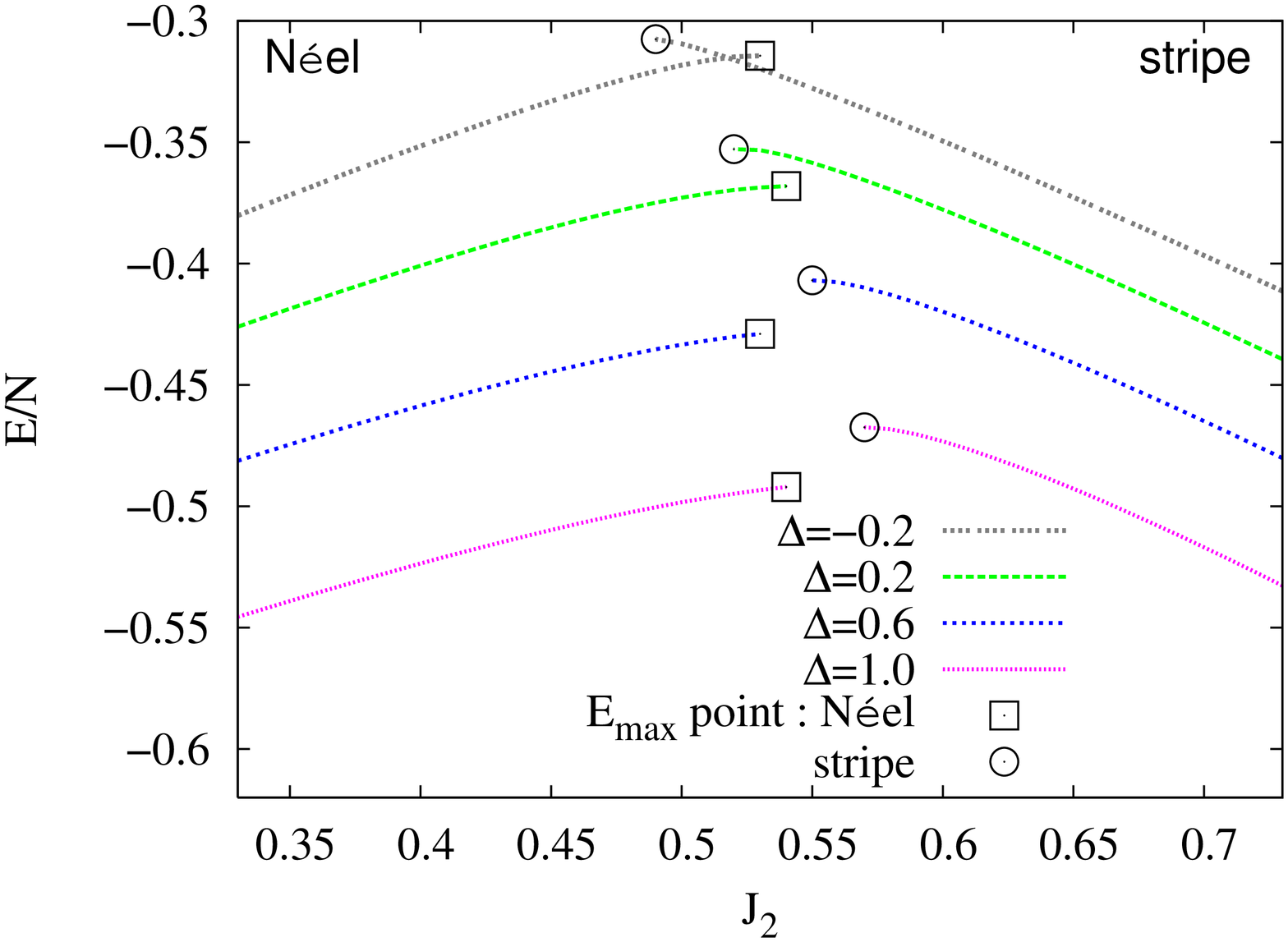,angle=270}}}
}
\caption{Extrapolated CCM LSUB$n$ results using the $z$-aligned and planar 
$x$-aligned states for the gs energy per spin, $E/N$, for the N\'{e}el and stripe 
phases of the $s=1/2$ $J_{1}^{XXZ}$--$J_{2}^{XXZ}$ model.  The LSUB$n$ results 
are extrapolated in the limit $n \rightarrow \infty$ using the sets $n=\{4,6,8,10\}$ 
for both the $z$-aligned states and the planar $x$-aligned states.  The NN exchange 
coupling $J_{1}=1$.  The meaning of the $E_{\mathrm{max}}$ points shown is described in the text.}
\label{E}
\end{center}
\end{figure*}
shows the extrapolated results for the gs energy per spin as a function of $J_{2}$ (with $J_{1}=1$) 
for various values of $\Delta$, for the $z$-aligned and planar $x$-aligned model states.  
For each model state, two sets of curves are shown, one (for smaller values of $J_{2}$) 
using the N\'{e}el state, and the other (for larger values of $J_{2}$) using the stripe 
state.  As we have discussed in detail elsewhere,\cite{Bi:1998,Ze:1998,Fa:2004} 
the coupled sets of LSUB$n$ equations (\ref{ket_coeff}) have natural termination points 
(at least for values $n > 2$) for some critical value of a control parameter 
(here the anisotropy, $\Delta$), beyond which no real solutions to the equations exist.  
The extrapolation of such LSUB$n$ termination points for fixed values of $\Delta$ to 
the $n \rightarrow \infty$ limit can sometimes be used as a method to calculate the 
physical phase boundary for the phase with ordering described by the CCM model state being 
used.  However, since other methods exist to define the phase transition points, 
which are usually more precise and more robust for extrapolation (as we discuss below), 
we have not attempted such an analysis here.  

Instead, in Fig.\ \ref{E}, the $E_{\mathrm{max}}$ points shown, for each set of calculations 
based on one of the four CCM model states used, are either those natural termination points 
described above for the highest (LSUB10) level of approximation we have implemented, or the 
points where the gs energy becomes a maximum should the latter occur first (i.e., as one 
approaches the termination point).  The advantage of this usage of the 
$E_{\mathrm{max}}$ points is that we do not 
then display gs energy data in any appreciable regimes where LSUB$n$ calculations with 
very large values of $n$ (higher than can feasibly be implemented) would not have 
solutions, by dint of having terminated already.

Curves such as those shown in Fig.\ \ref{E_Zaligned} illustrate very clearly that the 
corresponding pairs of gs energy curves for the $z$-aligned N\'{e}el and stripe phases 
cross one another for all values of $\Delta$ above some critical value, 
$\Delta \gtrsim 2.1$.  The crossings occur with a clear discontinuity in slope, as 
is completely characteristic of a first-order phase transition, exactly as observed in 
the classical (i.e., $s \rightarrow \infty$) case.  Furthermore, the direct first-order 
phase transition between the $z$-aligned N\'{e}el and stripe phases that is thereby 
indicated for all values of $\Delta \gtrsim 2.1$, occurs (for all such values of $\Delta$) 
very close to the classical phase boundary $J_{2}=\frac{1}{2}$, the point of 
maximum (classical) frustration.  Conversely, curves such as those shown in 
Fig.\ \ref{E_Zaligned} for values of $\Delta$ in the range $1 < \Delta \lesssim 2.1$ 
also illustrate clearly that the corresponding pairs of gs energy curves for the 
$z$-aligned N\'{e}el and stripe phases do not intersect one another.  In this regime we 
thus have clear preliminary evidence for the opening up of an intermediate phase between 
the N\'{e}el and stripe phases.  The corresponding curves in Fig.\ \ref{E_planar} 
for values of $\Delta < 1$ tell a similar story, with an intermediate phase similarly  
indicated to exist between the $xy$-planar-aligned N\'{e}el and stripe phases for 
values of $\Delta$ in the range $-0.1 \lesssim \Delta < 1$.

We show in Fig.\ \ref{M}
\begin{figure*}
\begin{center}
\mbox{
  \subfigure[$z$-aligned states]{\label{M_Zaligned}\scalebox{0.3}{\epsfig{file=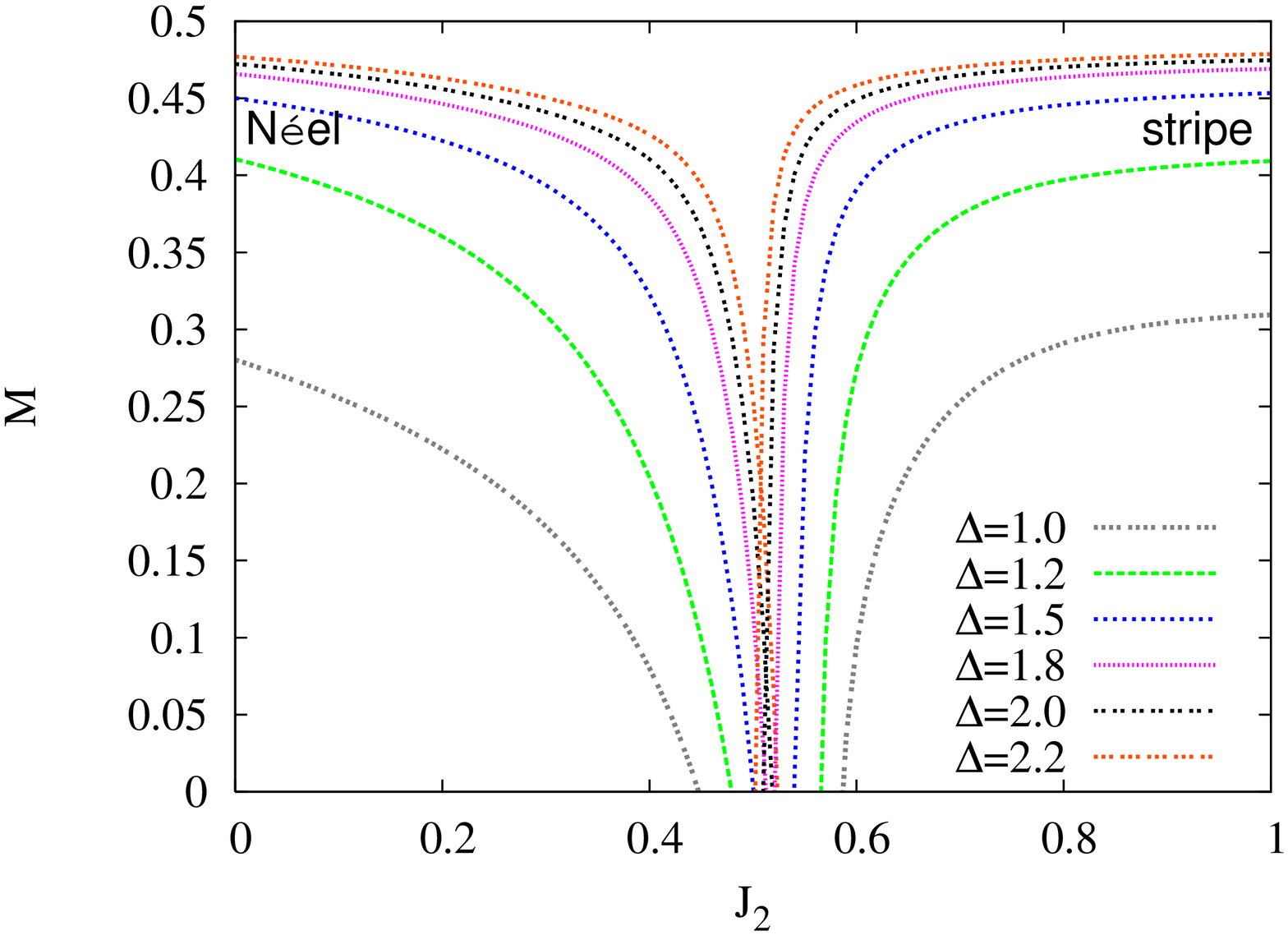,angle=270}}}
  \subfigure[planar $x$-aligned states]{\label{M_planar}\scalebox{0.3}{\epsfig{file=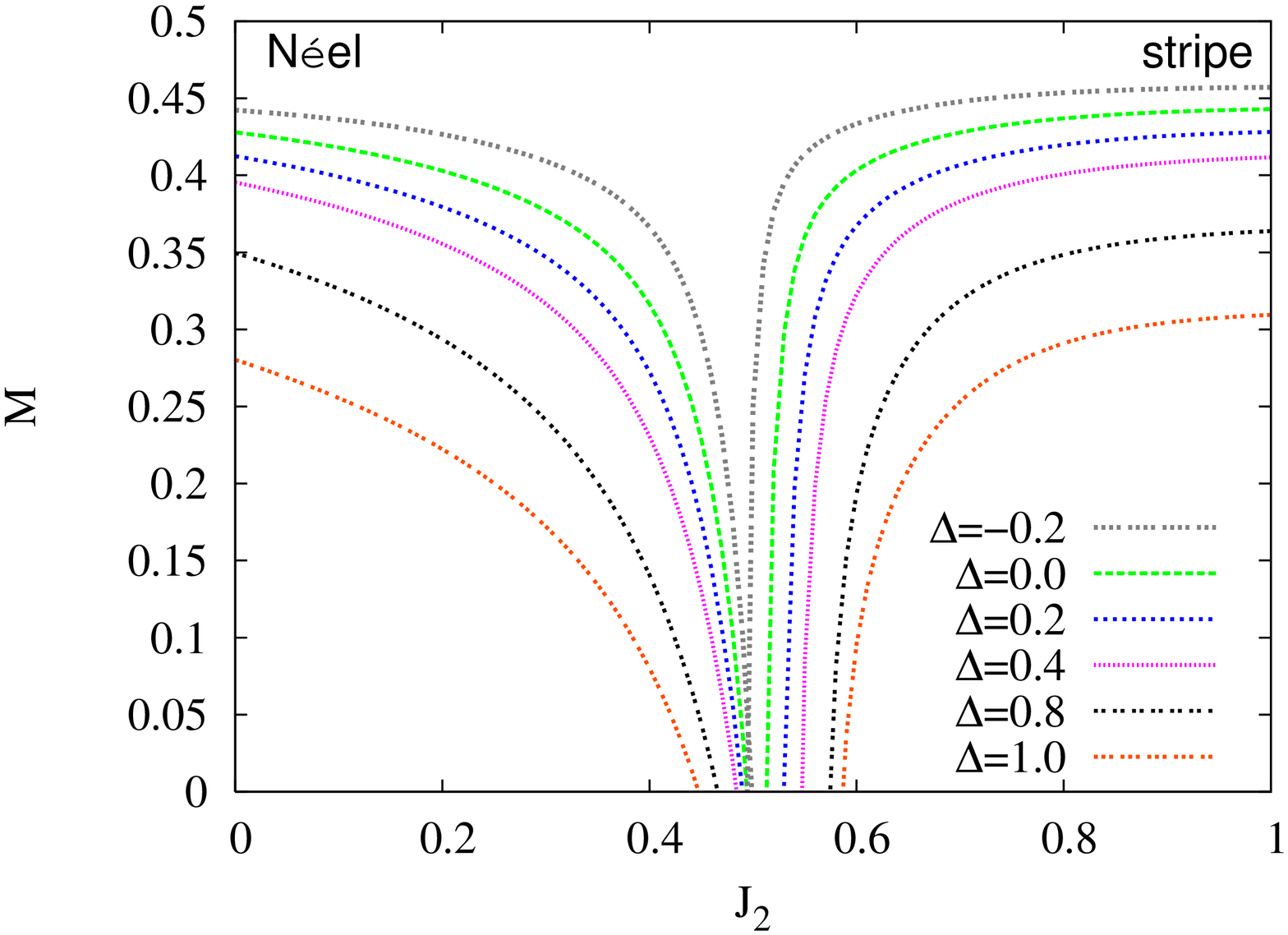,angle=270}}}
}
\caption{Extrapolated CCM LSUB$n$ results using the $z$-aligned and planar $x$-aligned 
states for the gs staggered magnetization, $M$, for the N\'{e}el and stripe phases of 
the $s=1/2$ $J_{1}^{XXZ}$--$J_{2}^{XXZ}$ model.  The LSUB$n$ results are extrapolated in 
the limit $n \rightarrow \infty$ using the sets $n=\{4,6,8,10\}$  for both the $z$-aligned 
states and the planar $x$-aligned states.  The NN exchange coupling $J_{1}=1$.}
\label{M}
\end{center}
\end{figure*}
corresponding indicative sets of CCM results, based on the same four model states, for the 
gs order parameter (viz., the staggered magnetization), to those shown in Fig.\ \ref{E} for 
the gs energy.  The staggered magnetization data completely reinforce the phase structure 
of the model as deduced above from the gs energy data.  Thus, let us now denote by $M_{c}$ 
the quantum phase transition point deduced from curves such as those shown in Fig.\ \ref{M}, 
where $M_{c}$ is defined to be either (a) the point where corresponding pairs of CCM staggered 
magnetization curves (for the same value of $\Delta$), based on the N\'{e}el and stripe 
model states, intersect one another if they do so at a physical value $M \geq 0$; 
or (b) if they do not so intersect at a value $M \geq 0$, the two points where the 
corresponding values of the staggered magnetization go to zero.  

Clearly, case (a) here corresponds to a direct phase transition between the N\'{e}el and 
stripe phases, which will generally be first-order if the intersection point has a value 
$M \neq 0$ (and, exceptionally, second-order, if the crossing occurs exactly at $M = 0$).  
On the other hand, case (b) corresponds to the situation where the points where the LRO 
vanishes for both quasiclassical (i.e., N\'{e}el-ordered and stripe-ordered) phases are, 
at least naively, indicative of a second-order phase transition from each of these phases 
to some unknown intermediate magnetically-disordered phase.  We return to a discussion 
of the actual order of such transitions in Sec.\ \ref{discussion}. In summary, we hence 
define the staggered magnetization criterion for a quantum critical point as 
the point where there is an indication of a phase transition 
between the two states by their order parameters becoming equal, 
or where the order parameter vanishes, whichever occurs first.  A detailed discussion 
of this order parameter criterion and its relation to the stricter energy 
crossing criterion may be found elsewhere.\cite{Schm:2006}  

From curves such as those shown in Fig.\ \ref{M_Zaligned}
we see that for $\Delta \lesssim 1.95$ for the 
$z$-aligned states, there exists an intermediate region between the critical points 
at which $M \rightarrow 0$ for the N\'{e}el and stripe phases.  Conversely, for 
$\Delta \gtrsim 1.95$ the two curves for the order parameters $M$ of the 
quantum N\'{e}el and stripe phases for the same value of $\Delta$ meet at a 
finite value, $M > 0$, as is typical of a first-order transition.  
Similarly, Fig.\ \ref{M_planar} shows that for the planar $x$-aligned states, there exists an 
intermediate region between the critical points at which $M \rightarrow 0$ for the 
N\'{e}el and stripe phases for all values of $\Delta$ in the range $-0.15 \lesssim \Delta < 1$.  
Again, the two curves for the order parameters $M$ of the N\'{e}el and stripe 
phases for the same value of $\Delta$ intersect at a value 
$M > 0$ for $\Delta \lesssim -0.15$.  In order to show more explicitly how the quantum 
phase transitions are driven by anisotropy, $\Delta$, we display the same data for 
the extrapolated results for the order parameter, $M$, somewhat differently in Fig.\ \ref{delta}, 
\begin{figure}[t]
\begin{center}
\epsfig{file=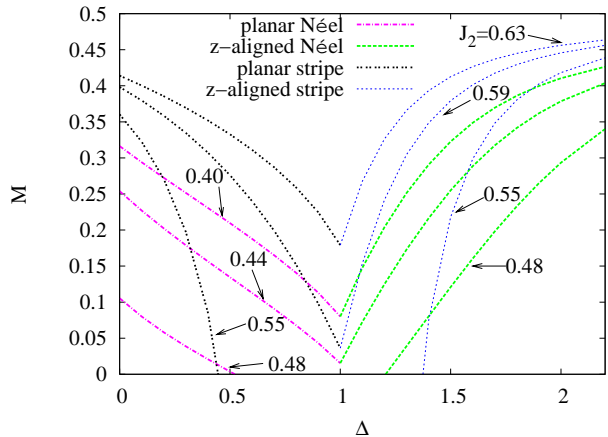,width=6cm,angle=270}
\caption{Extrapolated CCM LSUB$n$ results using the $z$-aligned and planar $x$-aligned 
states for the staggered magnetization versus the anisotropy $\Delta$ for the 
$s=1/2$ $J_{1}^{XXZ}$--$J_{2}^{XXZ}$ model, for the NN exchange coupling $J_{1}=1$.  
The LSUB$n$ results are extrapolated in the limit $n \rightarrow \infty$ using the sets 
$n = \{4,6,8,10\}$ for both the $z$-aligned model states and the planar 
$x$-aligned model states.}
\label{delta}
\end{center}
\end{figure}
where we plot $M$ as a function of $\Delta$ for various values of $J_{2}$ around the 
value $J_{2} = 0.5$, corresponding to the point of maximum (classical) frustration.

By putting together data of the sort shown in Figs.\ \ref{E}, \ref{M}, and 
\ref{delta} we are able to 
deduce the gs phase diagram of our 2D spin-1/2 $J_{1}^{XXZ}$--$J_{2}^{XXZ}$ model on 
the square lattice, from our CCM calculations based on the four model states with 
quasiclassical antiferromagnetic LRO (viz., the N\'{e}el and stripe states for both 
the $z$-aligned and planar $xy$-aligned cases).  We show in Fig.\ \ref{phase} 
\begin{figure}[t]
\begin{center}
\epsfig{file=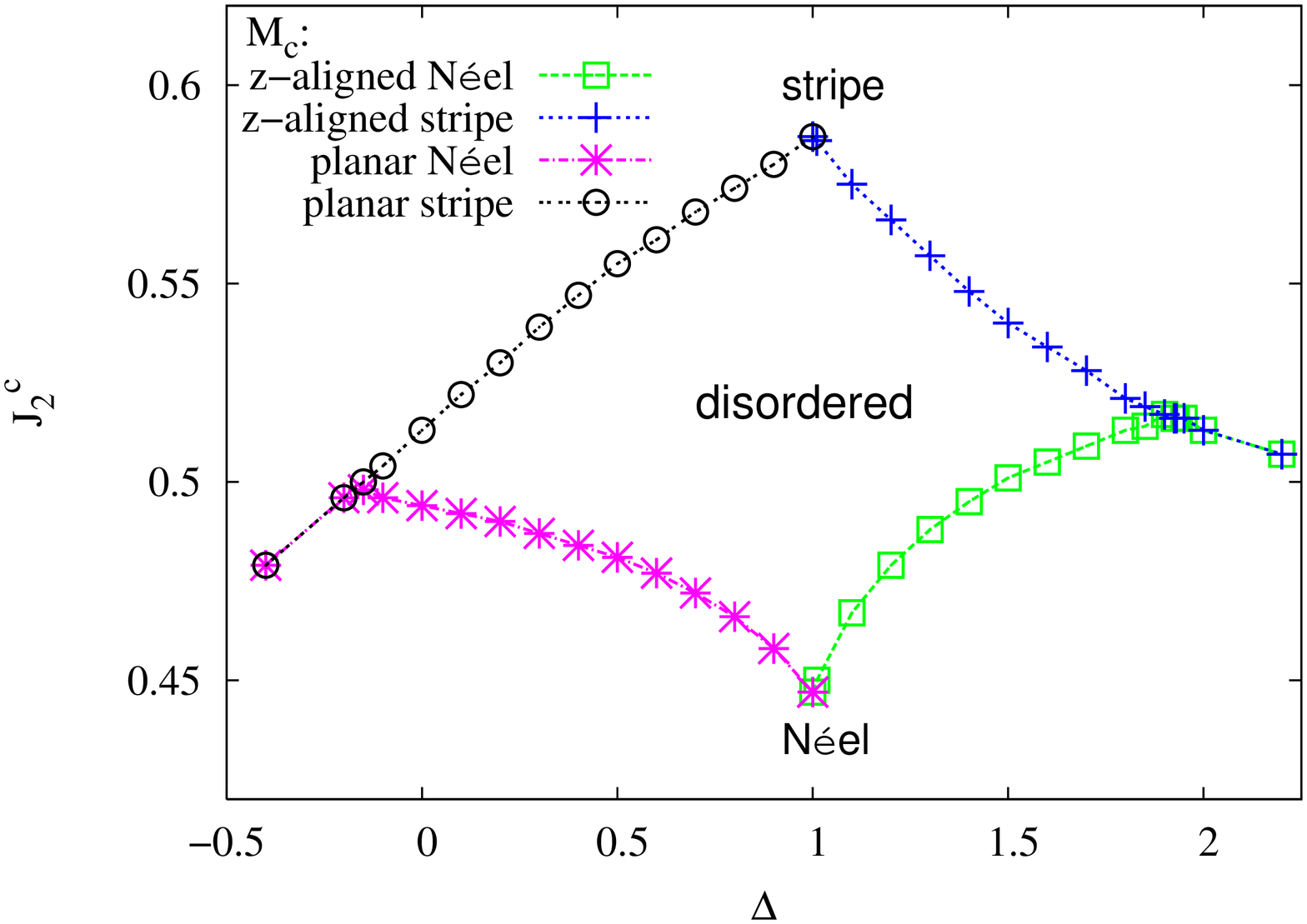,width=6cm,angle=270}
\caption{Extrapolated CCM LSUB$n$ results using the $z$-aligned and planar $x$-aligned 
states for the ground-state phase diagram of the $s=1/2$ $J_{1}^{XXZ}$--$J_{2}^{XXZ}$ model, 
for the NN exchange coupling $J_{1}=1$.  The LSUB$n$ results for the 
staggered magnetization are extrapolated to the 
limit $n \rightarrow \infty$ using the sets $n = \{4,6,8,10\}$ for both the $z$-aligned 
model states and the planar $x$-aligned model states.  $M_{c} \equiv $ 
magnetization critical point, defined in the text.}
\label{phase}
\end{center}
\end{figure}
the zero-temperature gs phase diagram, as deduced from the order parameter criterion,  
and using our extrapolated LSUB$n$ data sets with $n=\{4,6,8,10\}$, shown as the critical 
value $J_{2}^{c}$ for the NNN exchange coupling $J_{2}$ as a function of anisotropy $\Delta$ 
(with NN exchange coupling strength $J_{1}=1$).  Very similar results are obtained from 
using the energy criterion, where it can be applied (viz., along the transition lines 
between quasiclassical states with magnetic LRO).  In order to test the accuracy 
of our results, particularly the positions of the phase boundaries shown in 
Fig.\ \ref{phase}, we have also performed extrapolations using the LSUB$n$ data sets 
with $n=\{2,4,6,8,10\}$ and  $n=\{6,8,10\}$ for both the energy criterion and the 
order parameter criterion.  In general terms we find that the results are remarkably 
robust, and the error bars quoted below are based on such an analysis.

For the case of the $z$-aligned states, all of our results provide 
clear and consistent evidence for an upper {\it quantum triple point} (QTP) at 
($\Delta ^{c} = 2.05 \pm 0.15, J_{2}^{c} = 0.530 \pm 0.015$) (for $J_{1}=1$).  
For $1 < \Delta \lesssim 2.0$, there exists an intermediate paramagnetic 
(magnetically-disordered) quantum phase, separating the N\'{e}el and stripe phases.  
This intermediate phase disappears  for $\Delta \gtrsim 2.0$, and both our energy and 
order parameter criteria give clear and unequivocal evidence for a direct first-order 
quantum phase transition between the two quasiclassical antiferromagnetic states 
in this regime, just as in the corresponding classical model (i.e., with 
$s \rightarrow \infty$).  The phase boundary approaches the classical line 
$J_{2}^{c} = 0.5$ as $\Delta \rightarrow \infty$.

Similarly, for the case of the $xy$-planar-aligned phases, a second (lower) QTP occurs at 
($\Delta ^{c} = -0.10 \pm 0.15, J_{2}^{c} = 0.505 \pm 0.015$) (for $J_{1}=1$), with an 
intermediate disordered phase existing in the region $-0.1 \lesssim \Delta < 1$.  The 
$z$-aligned and $xy$-planar-aligned phases meet precisely at $\Delta = 1$, just as 
in the classical case.  Exactly at the isotropic point $\Delta = 1$, where the model 
becomes just the original $J_{1}$--$J_{2}$ model, the disordered phase exists for the 
largest range of values of $J_{2}$, $J_{2}^{c_{1}} < J_{2} < J_{2}^{c_{2}}$, as can 
be clearly seen from Fig.\ \ref{phase}.  For the pure $J_{1}$--$J_{2}$ model our 
calculations yield the values $J_{2}^{c_{1}}/J_{1} = 0.44 \pm 0.01$ and 
$J_{2}^{c_{2}}/J_{1} = 0.59 \pm 0.01$ that demarcate the phase boundaries for 
the disordered phase, in complete agreement with both our own earlier work and that 
of others that we have already discussed in Sec.\ \ref{introd}.

\section{Discussion and Conclusions}
\label{discussion}
We have shown in detail how, as expected, the quantum fluctuations present in the 
spin-1/2 $J_{1}$--$J_{2}$ model on the 2D square lattice, that has become an archetypal 
model for studying the interplay between quantum fluctuations and frustration, 
can be tuned by the introduction of spin anisotropy.  We have clearly confirmed our 
prior expectation that anisotropy reduces the quantum fluctuations.  Thus, 
for both the cases $\Delta > 1$ and $0 < \Delta < 1$, the intermediate 
paramagnetic phase present in the pure $J_{1}$--$J_{2}$ model is observed 
to shrink to a smaller range of values of $J_{2}/J_{1}$ centered near to the 
point of maximal classical frustration, $J_{2}^{c}/J_{1}=\frac{1}{2}$, that 
marks the classical phase boundary between the N\'{e}el-ordered and 
collinear stripe-ordered phases.

We have seen that the intermediate disordered phase disappears precisely at two 
quantum triple points at $\Delta^{c} = -0.10 \pm 0.15$ and $\Delta^{c} = 2.05 \pm 0.15$, 
and that for values of $\Delta$ outside the range spanned by these values the 
intermediate phase is totally absent.  In particular, for $\Delta \gtrsim 2.0$ we 
find unequivocal evidence for a first-order phase transition between the N\'{e}el 
and collinear stripe phases.  This direct first-order phase transition between states 
of different quasiclassical antiferromagnetic ordering is very similar to what 
has been observed in another similar extension of the spin-1/2 $J_{1}$--$J_{2}$ 
model on a square lattice, namely the so-called $J_{1}$--$J_{2}$--$J_{\perp}$ model 
on a stacked square lattice where we now introduce a (weak) interlayer 
coupling through NN bonds of strength $J_{\perp}$.  The quantum fluctuations in the 
$J_{1}$--$J_{2}$ model are tuned here by the parameter $J_{\perp}$.  An analysis of 
this model\cite{Schm:2006} found that the intermediate region of disordered paramagnetic 
phase, $\alpha^{c_1}<J_{2}/J_{1}<\alpha^{c_2}$, in the pure $J_{1}$--$J_{2}$ model now 
shrinks as the interlayer coupling strength $J_{\perp}$ is increased.  The second-order 
phase transition for the N\'{e}el-ordered phase to the paramagnetic phase disappears for 
$J_{\perp}/J_{1}$ above some critical value (estimated to be in the range 0.2--0.3) 
marking  a QTP in the $J_{2}$--$J_{\perp}$ plane (with $J_{1} \equiv 1$).  Above the 
QTP there is again a direct first-order phase transition between the two phases of 
different quasiclassical antiferromagnetic LRO.

On the other hand this scenario of a first-order phase transition between the two states 
of different quasiclassical LRO may be contrasted with the situation observed in yet 
another generalization of the pure spin-1/2 $J_{1}$--$J_{2}$ model on a square lattice, 
namely the so-called $J_{1}$--$J_{1}'$--$J_{2}$ model that we have briefly discussed 
in Sec.\ \ref{introd}.   In this case the quantum fluctuations are tuned by introducing 
a spatial anisotropy so that the NN bonds have different strengths in the intrachain 
($J_{1}$) and interchain ($J_{1}'$) directions on the square lattice.  A similar CCM 
analysis of the spin-1/2 version of this model by some of 
the present authors\cite{Bi:2007_j1j2j3_spinHalf} again 
found a QTP in the $J_{2}$--$J_{1}'$ plane (with $J_{1} \equiv 1$), now below which 
the disordered paramagnetic phase disappears, and there is again a direct phase transition 
between the quasiclassical N\'{e}el and stripe-ordered phases with magnetic
LRO.  However, the surprising and novel situation found here was the existence of strong 
evidence for the phase transition in this case to be second-order, and hence inexplicable 
by standard Ginzburg-Landau theory, as discussed more fully in Sec.\ \ref{introd} above.   

Having discussed the transition line between the two phases of quasiclassical 
antiferromagnetic LRO in the phase diagram in the $J_{2}$--$\Delta$ plane 
of our spin-1/2 $J_{1}^{XXZ}$--$J_{2}^{XXZ}$ model on the 2D square lattice, we turn 
our attention to the four phase boundary lines shown in Fig.\ \ref{phase} 
that delimit the region of existence for the intermediate disordered 
paramagnetic phase.  As has been explained in Ref.\ [{\onlinecite{Schm:2006}] a judicious 
combination of the CCM energy data with the CCM order parameter data can shed light on 
the nature of the phase transitions between the quasiclassically long-range-ordered phases 
and the paramagnetic phase.  The method for so doing relies essentially on the fact 
that although we perform our CCM calculations with model (or reference) states with 
quasiclassical LRO, one knows\cite{Bi:1998_PRB58,Ze:1998,Kr:2000,Dar:2005} that one can 
also reliably use such calculations in parameter regimes where all semblance of the 
quasiclassical LRO is destroyed.  Thus, what is required for the CCM equations to 
converge to a solution is a sufficient overlap between the wave functions of the 
model (reference) state $|\Phi\rangle$ and the true GS $|\Psi\rangle$.  The 
termination points of the CCM LSUB$n$ equations discussed above are indicators of 
where this condition breaks down.  Thus, provided that the CCM LSUB$n$ equations 
converge and yield extrapolated solutions far enough beyond the points $M_{c}$ where 
the order parameter vanishes, we can also determine whether the solution based on the 
N\'{e}el-ordered or the stripe-ordered model state has lower energy.  

We find in this way 
that there are indicators of a very narrow region where the gs energy obtained with the 
N\'{e}el model state might be slightly lower in energy than that obtained with the 
collinear striped model state, even in regions (close to) where the N\'{e}el order 
parameter has already gone to zero, but where the stripe order parameter is still 
nonzero.  As explained in more detail in Ref.\ [{\onlinecite{Schm:2006}], 
the use of this evidence here 
points towards the zero-temperature phase transitions from N\'{e}el LRO to quantum 
paramagnetic disorder being second-order, while the transitions from quantum 
paramagnetic disorder to collinear stripe order are possibly (rather weakly) 
first-order rather than second-order.  We stress, however, that the analysis here is 
very sensitive to the accuracy of our results, and the evidence for the nature 
of these quantum phase transitions involving the quantum paramagentic state 
in the regime $-0.1 \lesssim \Delta \lesssim 2.0$ is less compelling 
than that for the transition between the two quasiclassically ordered states 
being first-order in the regime $\Delta \gtrsim 2.0$.

The only other analysis of the current spin-1/2 $J_{1}^{XXZ}$--$J_{2}^{XXZ}$ model 
on the square lattice of which we are aware\cite{Be:1998} has been performed at the very 
low level of lowest-order spin-wave theory (LSWT).  For the case studied here of 
equal spin-anisotropy parameters in the NN and NNN exchange bonds, these authors have 
only investigated the case $\Delta > 1$, for which they find an (upper) QTP at 
a very small value of the anisotropy parameter, $\Delta^{c}_{(u)} \approx 1.048$, 
much smaller than the corresponding value $\Delta^{c}_{(u)} = 2.05 \pm 0.15$ 
obtained by us for the upper QTP.  Such an extreme fragility or sensitivity of the 
paramagnetic phase to spin anisotropy is not easy to understand.  In the face of 
our own much more accurate calculations it would seem simply to be an artefact of 
the LSWT approximation.  On the other hand, the LSWT analysis does give the same 
qualitative trends as found by us for the phase transitions in the range $\Delta > 1$, viz., 
a second-order transition between the N\'{e}el-ordered and disordered phases, and 
a first-order transition between the disordered and collinear stripe-ordered phases 
for $1 < \Delta < \Delta^{c}_{(u)}$, and a direct first-order transition between the 
N\'{e}el-ordered and collinear stripe-ordered phases for $\Delta > \Delta^{c}_{(u)}$.  

It is perhaps worth noting at this point in the context of spin-wave theory (SWT) that 
Igarashi\cite{Ig:1993} has shown that whereas its lowest-order (or linear) version 
(LSWT) works quite well when applied to the isotropic Heisenberg model with NN couplings only, 
it consistently oversestimates the quantum fluctuations in the pure (isotropic) 
$J_1$--$J_2$ model as the frustration $J_{2}/J_{1}$ increases.  Thus, he showed by 
going to higher orders in SWT in powers of $1/s$, where LSWT is the leading order, 
that the expansion converges reasonably well for values of $\alpha \equiv J_{2}/J_{1} \lesssim 0.35$, 
but for larger values of the frustration parameter $\alpha$, including the point $\alpha = 0.5$ of 
maximum classical frustration, the series loses stability.  He showed for the $s = \frac{1}{2}$ 
$J_{1}$--$J_{2}$ model that whereas LSWT predicts\cite{Ch:1988}  
a value of $\alpha^{c_{1}} \approx 0.38$ at which 
the transition from the N\'{e}el-ordered phase to the disordered phase occurs, the 
higher-order corrections to SWT for $\alpha \lesssim 0.4$ make the N\'{e}el-ordered phase 
more stable than predicted by LSWT.  This is precisely in agreement with our own 
predicted value of $\alpha^{c_{1}} = 0.44 \pm 0.01$ for the $s = \frac{1}{2}$ 
$J_{1}$--$J_{2}$ model on the square lattice.  He concludes that any predictions from 
SWT for the $J_{1}$--$J_{2}$ model on the square lattice are likely to be unreliable for 
values $J_{2}/J_{1} \gtrsim 0.4$.

For reasons unclear to us, the authors of Ref.\ [\onlinecite{Be:1998}] never 
investigated the regime with $\Delta < 1$, for which we find a lower QTP at 
$\Delta^{c}_{(l)} = -0.10 \pm 0.15$, $J_{2(l)}^{c} = 0.505 \pm 0.015$.  Clearly, 
our results are consistent with this lower QTP occurring exactly at the isotropic 
{\it XY} point (i.e., $\Delta = 0$) of the model, and also exactly at the point of 
maximal classical frustration, $J_{2} = \frac{1}{2}$.  A more detailed theoretical 
investigation of the corresponding $J_{1}^{XX}$--$J_{2}^{XX}$ model is clearly 
warranted by our results.  

Finally, we note that in our analysis here we have relied on
two of the unique strengths of the CCM, namely its ability
to deal with highly frustrated systems as readily as unfrustrated
ones, and its use from the outset of infinite lattices.
In turn, these features lead to its ability to yield accurate phase
boundaries even in the very delicate regions near quantum triple points.
Our own results for the gs energy and staggered magnetization from 
four sets of independent calculations based on different reference states 
provide us with a set of internal checks that lead us
to believe that we now have a self-consistent and robust 
description of this rather challenging model system.

\section*{ACKNOWLEDGMENTS}
Two of us (RFB and PHYL) are grateful to Professor C.E. Campbell for useful 
discussions and to the University of Minnesota Supercomputing 
Institute for Digital Simulation and Advanced Computation  
for the grant of supercomputing facilities in conducting this research.  
We also thank Stephan Mertens and his group at the University of Magdeburg
for giving us computing time on their Beowulf cluster Tina.  Two of us (RD and JR) 
are grateful to the DFG for support (through project Ri615/16-1).

\end{document}